\documentclass[12pt]{article}
\usepackage{epic}
\def\Pa{Painlev\'e }
\def\dPf{\mbox{d-P${}_{\rm IV}$}}
\def\Pf{\mbox{P${}_{\rm IV}$}}
\def\Pt{\mbox{P$_{\rm II}$}}
\newcommand{\tfrac}[2]{\textstyle{\frac{#1}{#2}}}
\newcommand{\osa}[3]{${\frac{[\,#1\,]}{[\,#2\,][\,#3\,]}}$}
 
\begin{document}
 
\title{Rational solutions to d-P${}_{\rm IV}$}
\author{Jarmo Hietarinta${}^{1,2}$\thanks{E-mail:
hietarin@newton.tfy.utu.fi}  
\phantom{$\!\!\frac1{\frac11}\!$}
and
Kenji Kajiwara${}^{2}$\thanks{E-mail: kaji@elrond.doshisha.ac.jp}\\
${}^{1}$Department of Physics, University of Turku\\FIN-20014 Turku,
Finland\\
${}^{2}$Department of Electrical Engineering\\Doshisha University,
Tanabe, Kyoto 610-03, Japan}
\date{}
\maketitle

\begin{abstract}
We study the rational solutions of the discrete version of
Painlev\'e's fourth equation (\dPf). The solutions are generated by
applying Schlesinger transformations on the seed solutions $-2z$ and
$-1/z$. After studying the structure of these solutions we are able to
write them in a determinantal form that includes an interesting
parameter shift that vanishes in the continuous limit.
\end{abstract}

\section{Introduction}
One important question in the study of discrete versions of continuous
differential equations concerns the existence of corresponding special
solutions. For continuous \Pa equations rational and special function
solutions are known \cite{Luka,Oka}, and in many cases even a rigorous
classification has been done \cite{Murata}.  If one proposes a
discrete version of a \Pa equation it is not enough that in some
continuous limit the continuous \Pa equation is obtained, but in
addition the proposed equation should share some further properties of
the original equation. One of these properties should be the
equivalent of the \Pa property, called ``singularity confinement''
\cite{GRP}. This has already been used to propose discrete forms of
the \Pa equations \cite{RGH}. Other structures of the continuous \Pa
equations that have been shown to exist for the discrete ones include
their relationships by coalescence limits \cite{coales} and the
existence of Hirota forms \cite{bi} for these equations. What is still
largely an open question is the fate of the special solutions
(rational, algebraic, special function) known for the continuous case.

Here we discuss the determinantal structure of the {\em rational
solutions} to the discrete fourth Painlev\'e equation, (\dPf), given
by \cite{RGH}
\begin{eqnarray}
\lefteqn{(x_{n+1}+x_n)(x_n+x_{n-1})=
\phantom{|_|}}\nonumber\\ &&\ \hskip 2cm
\frac{(x_n+\alpha+\beta)(x_n+\alpha-\beta)
(x_n-\alpha+\beta)(x_n-\alpha-\beta)}{(x_n+z_n+\gamma)(x_n+z_n-\gamma)}
\label{E:dP4}
\end{eqnarray}
where $z_n=\delta n+\zeta$.  [Note that this equation is invariant under
the change sign of any of parameters $\alpha,\, \beta,\,\gamma$.]

One reason for calling (\ref{E:dP4}) \dPf\ is that if we put $\alpha^2
= -a + \sqrt{-b/2} + \delta^{-2},\, \beta^2 = -a - \sqrt{-b/2} +
\delta^{-2},\, \gamma = \delta^{-1}$ and then take the limit
$\delta\to0$ we get \Pf:
\begin{equation}
\frac{d^2w}{dz^2}=\frac{1}{2w}\left(\frac{dw}{dz}\right)^2
+\frac{3}{2}w^3+4zw^2+2(z^2-a)w+\frac{b}{w}\ .
\label{P4}
\end{equation}
The comparison of (\ref{E:dP4}) and (\ref{P4}) reveals the first
interesting difference between the discrete and continuous
versions: the discrete one has more parameters.

But why should one insist on writing the solution in determinantal
form? It is well known that most integrable systems have multisoliton
(and rational) solutions in determinantal form, and the same holds for
many special solutions of continuous Painlev\'e equations.  This
determinantal structure actually carries fundamental information and
reveals the ``basic'' object hidden in the solutions.  For example it
is well known that \Pt ,
\begin{equation}
\frac{d^2u}{dz^2}=2u^3-4zu + 4\alpha,
\end{equation}
where $\alpha$ is a parameter, admits rational solutions for
$\alpha=N+1$. It has also been shown that these solutions can be
are expressed as \cite{Vorobev}
\begin{equation}
u=\frac{d}{dz}\log \frac{\tau_{N+1}}{\tau_N},
\label{E:taup2}
\end{equation}
where $\tau_N$'s are polynomials in $z$ (Vorob'ev--Yablonski
polynomials).  However, it has been shown only recently that $\tau_N$
can be expressed in determinantal form \cite{P2},
\begin{equation}
\tau_N=\left|\matrix{ q_{N}(z,t) & q_{N+1}(z,t) & \cdots &
q_{2N-1}(z,t)\cr q_{N-2}(z,t) & q_{N-1}(z,t) & \cdots &
q_{2N-3}(z,t)\cr \vdots & \vdots & \ddots & \vdots \cr q_{-N+2}(z,t) &
q_{-N+3}(z,t) & \cdots & q_{1}(z,t)\cr}\right|\ , 
\label{P2tau}
\end{equation}
where $q_k$'s are the so called Devisme polynomials defined by
\begin{equation}
\sum_{k=0}^\infty q_k(z,t)\lambda^k=
{\rm exp}\left( z\lambda + t\lambda^2+\frac{1}{3}\lambda^3\right),
\quad {\rm and }\ \ q_k(z,t)=0,\ {\rm for}\ k<0\ .\label{gen}
\end{equation} 
Thus, these Devisme polynomials may be considered as basic polynomials
in the rational solutions of \Pt. Similar determinantal structure of
the rational solutions is observed for the discrete case: in fact, it has
been shown that in the case of d-\Pt, the rational solutions can be
expressed by determinants whose entries are given by Laguerre
polynomials \cite{dP2-rational_1,dP2-rational_2}.

In the case of \Pf, it is well known \cite{Luka} that the continuous
\Pf\ has three rational solution hierarchies, whose ``seed'' solutions
are
\begin{equation}
y(z)=-2z,\quad-\frac1z,\mbox{ or }-\frac23z,
\end{equation}
and determinantal forms of the first two hierarchies have been
discussed in general terms, e.g., in \cite{Oka}.  The detailed results
for the first two hierarchies are as follows \cite{KOu}. Let
$\tau_N^\nu$ be an $N\times N$ determinant of Hankel type given by
\begin{equation}
\tau_N^\nu=\det \left| H_{\nu+i+j-2}\right|_{1\leq i,j\leq N}\ ,
\quad \tau_0^\nu=1\ ,
\label{P4tau}
\end{equation}
where $H_n$, $n=0,1,2\cdots$ are Hermite polynomials in $x$
characterized by the recursion relations
\begin{equation}
H_{n+1}=xH_n-nH_{n-1},\quad \frac{d}{dx}H_n=nH_{n-1},\quad H_0=1,\ H_1=x.
\label{Hermite}
\end{equation}
Then 
\begin{equation}
w=-\sqrt{2}\frac{d}{dx}\log\left(\frac{\tau_{N+1}^\nu}{\tau_N^\nu}\right)
\ ,\quad z=\frac{1}{\sqrt{2}} x\ ,\label{-1/z}
\end{equation}
are rational solutions of P$_{\rm IV}$(\ref{P4}) for parameter values
\begin{equation}
(a,b)=(-(\nu+2N+1),-2\nu^2),\quad\nu,N\in{\cal Z},\quad
\nu\geq 1,\quad N\geq 0.
\label{param1_kaji}
\end{equation}
Moreover, 
\begin{equation}
w=\sqrt{2}\left(
\frac{d}{dx}\log\left(\frac{\tau_{N+1}^\nu}{\tau_N^{\nu+1}}\right)
-x\right)\ ,\quad z=\frac{1}{\sqrt{2}} x\ ,\label{-2z}
\end{equation}
yield rational solutions of P$_{\rm IV}$(\ref{P4}) with
\begin{equation}
(a,b)=(N-\nu,-2(\nu+N+1)^2),\quad \nu,N\in{\cal Z},\quad
\nu\geq 0,\quad N\geq 0.
\label{param3_kaji}
\end{equation}
The solutions given in (\ref{-1/z}) and (\ref{-2z}) correspond to the 
``$-1/z$'' and ``$-2z$'' hierarchies, respectively\footnote{ Equations
(\ref{-1/z}) and (\ref{-2z}) actually give half of the solutions in
$-1/z$ and $-2z$ hierarchies obtained by Murata \cite{Murata}. The
other half is given in terms of polynomials similar to the Hermite
polynomials, but with some different signs in the recursion relations
above.}.

Here our object is to find the discrete versions of these results for
the $-2z$ and $-1/z$ hierarchies.  (We hope to return the more
complicated $-\frac23z$ case elsewhere.) We cannot simply discretize the
continuous results, because the discrete equation has more parameters,
and the way the new parameters modify the continuous results is indeed
one of the interesting questions.  In approaching this problem we will
not use any detailed properties of the continuous case, we just assume
that that such rational solutions should arise from determinants of
polynomials, which always happens in the continuous case.  Starting with
the discrete version of the usual seed solutions, we first construct a
set of solutions in both hierarchies using B\"acklund--Schlesinger
transformations (discussed in the next section) and then search for the
determinantal structure by studying the properties of these rational
solutions, the main clue being the factorization of the denominator.

\section{B\"acklund-Schlesinger transformation}
In order to generate rational solutions we use the
B\"acklund--Schlesinger transformations on the seed solutions. These
transformations were given in \cite{Schl}.  We write them as follows:

\vskip 0.2cm
\noindent
{\bf B\"acklund transformation:} Let us assume that $x(n)$ solves
\dPf\ with parameter values $(\alpha,\beta,\gamma)$ (in which case we
often write $x=x(n;\alpha,\beta,\gamma)$ or
$x=\{f(n);\alpha,\beta,\gamma\}$).  Using these
let us define
\begin{equation}
BT_{a,b}(x):=\left[\frac{x\bar x+\bar x(\tilde z+a)+ x(\tilde
z-a)+b^2-a^2)}{x+\bar x}\right]_{n\to n+\frac12}
\end{equation}
where $\tilde z=(n+\frac12)\delta+\zeta$, and $\bar x=x(n+1)$.  Note
the shifts in $n$.  Then we get new solutions from the elementary
B\"acklund transformations $BT_i$ as follows:
\begin{equation}\begin{array}{lll}
&&x(n;\alpha+\tfrac12\delta,\gamma,\beta)=
BT_1x(n)=BT_{\alpha,\beta}(x(n;\alpha,\beta,\gamma),\\
&&x(n;\alpha-\tfrac12\delta,\gamma,\beta)=
BT_2x(n)=BT_{-\alpha,\beta}(x(n;\alpha,\beta,\gamma),\\
&&x(n;\beta+\tfrac12\delta,\gamma,\alpha)=
BT_3x(n)=BT_{\beta,\alpha}(x(n;\alpha,\beta,\gamma),\\
&&x(n;\beta-\tfrac12\delta,\gamma,\alpha)=
BT_4x(n)=BT_{-\beta,\alpha}(x(n;\alpha,\beta,\gamma).
\end{array}
\end{equation}
These transformations jump too much in the parameter space and
therefore it is useful to define {\bf Schlesinger transformations}
that generate new solutions by changing only one of the parameters
\begin{equation}\begin{array}{lll}
&&x(n;\alpha+\delta,\beta,\gamma)=S_1(x)=
BT_1BT_1\,x(n;\alpha,\beta,\gamma),\\
&&x(n;\alpha-\delta,\beta,\gamma)=S_2(x)=
BT_2BT_2\,x(n;\alpha,\beta,\gamma),\\
&&x(n;\alpha,\beta+\delta,\gamma)=S_3(x)=
BT_2BT_3BT_1BT_3\,x(n;\alpha,\beta,\gamma),\\
&&x(n;\alpha,\beta-\delta,\gamma)=S_3(x)=
BT_2BT_3BT_2BT_4\,x(n;\alpha,\beta,\gamma).
\end{array}
\end{equation}

It should be noted that sometimes there are barriers over which
Schlesinger transformations cannot cross. For example if
$\alpha=\gamma+(m+\frac12)\delta$ and $\beta=\gamma-(n+\frac12)\delta$
(which is relevant for some rational hierarchies) then the barrier
$n=0$ cannot be crossed by Schlesinger transformations because $BT_3$
yields the 0 solution, similarly for $m=0$ with $BT_2$.

\section{Some solutions} 
Using the above Schlesinger transformations we can construct other
rational solutions from the seed solutions. The seed for the $-2z$
hierarchy is given by
\begin{equation}
x_{00}=\{ - 2 z,\gamma +
\tfrac12 \delta,\gamma - \tfrac12 \delta,\gamma\}.  
\end{equation}
(here and in the following $z=n\delta+\zeta$.)  Since equation
(\ref{E:dP4}) is invariant under the sign changes in $\alpha,\,\beta$
and $\gamma$ we may always assume the above signs in front of
$\gamma$.  With Schlesinger transformations we can reach the
parameter values
\begin{equation}
\alpha=\gamma+(M+\tfrac12)\delta,\quad
\beta=\gamma-(N+\tfrac12)\delta, \label{E:ab2} 
\end{equation} 
where $M$ and $N$ are nonnegative integers, and we may also assume
that $M\ge N $ because of the $\alpha\leftrightarrow \beta$
symmetry. This then defines the extent of the hierarchy. For
particular solutions we give $N,M$ as subscripts. Some further
solutions obtained this way are (see also \cite{CB})
\begin{eqnarray*} 
x_{10}&=& \left\{ \frac{ - 2 z^2 + \delta (\gamma + \delta)}{z},
\gamma + \tfrac32 \delta, \gamma - \tfrac12 \delta ,\gamma\right\},\\
x_{20}&=& \left\{ \frac{4 z^3 + 2 z \delta ( - 3 \gamma - 5 \delta)} {
- 2 z^2 + \delta (\gamma + 2 \delta)}, \gamma + \tfrac52 \delta,
\gamma - \tfrac12 \delta ,\gamma\right\},\\ x_{30}&=& \left\{ \frac{4
z^4 + 4 z^2 \delta ( - 3 \gamma - 7 \delta) + 3 \delta^2 (\gamma^2 + 5
\gamma \delta + 6 \delta^2)}{ - 2 z^3 + z \delta (3 \gamma + 8
\delta)}, \gamma + \tfrac72 \delta, \gamma - \tfrac12 \delta
,\gamma\right\},\\ x_{31}&=& \left\{ \frac{8 z^5 - 16 z^3 \delta^2 + 2
z \delta^2 (3 \gamma^2 + \delta^2)}{ - 4 z^4 + 4 z^2 \delta ^2 +
\delta^2 (\gamma^2 - \delta^2)}, \gamma + \tfrac32 \delta, \gamma -
\tfrac32 \delta ,\gamma\right\}.
\end{eqnarray*}

The $-1/z$-hierarchy is not connected to the $-2z$ hierarchy by a
Schlesinger transformation, but only by a B\"acklund transformation
followed by a redefinition of $\gamma$. In any case the seed is
\begin{equation}
x_{10}= \left\{ \frac{ - \delta(\gamma + \delta)}{z}, \gamma +
\tfrac32\delta, \gamma + \tfrac12\delta \right\},
\end{equation}
and in general the parameter values in this hierarchy are
\begin{equation}
\alpha=\gamma+(M+\tfrac12)\delta,\quad \beta=\gamma+(N+\tfrac12)\delta,
\label{E:ab1}
\end{equation}
with $0\le N<M$. Some further solutions are given by
\begin{eqnarray*}
x_{20}&=& \left\{ \frac{{2z\delta(2 \gamma + 3 \delta)}} { - 2 z^2 +
\delta(\gamma + 2 \delta)}, \gamma + \tfrac52 \delta, \gamma +
\tfrac12 \delta,\gamma \right\}
,\\
x_{21}&=& \left\{ \frac{{\delta(\gamma + 2 \delta)( - 2 z^2 +
\delta(\gamma + \delta))}} {2 z^3 + z \delta\gamma}, \gamma + \tfrac52
\delta, \gamma + \tfrac32 \delta,\gamma \right\}
,\\
x_{30}&=& \left\{ \frac{{3\delta( - \gamma - 2 \delta)( - 2 z^2 +
\delta(\gamma + 3 \delta))}} { - 2 z^3 + z\delta (3 \gamma + 8
\delta)}, \gamma + \tfrac72 \delta, \gamma + \tfrac12 \delta,\gamma
\right\}
,\\
x_{32}&=& \left\{ \frac{{ (- \gamma - 3 \delta)\delta(4 z^4 - 4 z^2
\delta^2 + 3 \delta^2(\gamma^2 + 2 \gamma \delta + \delta^2))}} {4 z^5
+ 4 z^3 \delta (2 \gamma + \delta) + z\delta^2 (3 \gamma^2 + 4 \gamma
\delta + \delta^2)}, \gamma + \tfrac72 \delta, \gamma + \tfrac52
\delta,\gamma\right\}.
\end{eqnarray*}
It is obvious from these examples (and from the equation) that there is
an overall scaling invariance, and because of this we will in the
following simplify expressions by scaling out $\delta$ by writing
$\gamma=c\delta,\, z=n\delta$.

\section{The elementary polynomials}
A common property for rational and other special function solutions of
the continuous case is that the denominator factorizes into two
determinants.  Thus after constructing a set of solutions (all 
computations were done using REDUCE \cite{RED}) we studied the
factorization of their denominators, the results are given in Figure 1. 
\begin{figure}[t]
\begin{picture}(200,245)(-100,-10)
\multiputlist(0,225)(40,0){ , , ,0,\osa634,3
,$\gamma+\tfrac72\delta$}
\multiputlist(0,200)(40,0){$-1/z$, ,0,\osa432,\osa946,2
,$\gamma+\tfrac52\delta$}
\multiputlist(0,175)(40,0){ ,0,\osa212,\osa524,\osa836,1
,$\gamma+\tfrac32\delta$}
\multiputlist(0,150)(40,0){0,\osa001,\osa102,\osa203,\osa304,0
,$\gamma+\tfrac12\delta$}
\multiputlist(-40,121)(40,0){$M=$,0,1,2,3,4}
\multiputlist(-40,112)(40,0){ , , , , , , N,$\beta$}
\multiputlist(-40,104)(40,0){$\alpha=$,$\gamma+\tfrac12\delta$,
$\gamma+\tfrac32\delta$,$\gamma+\tfrac52\delta$,$\gamma+\tfrac72\delta$,
$\gamma+\tfrac92\delta$}
\multiputlist(0,75)(40,0){0,\osa001,\osa102,\osa203,\osa304,0
,$\gamma-\tfrac12\delta$}
\multiputlist(0,50)(40,0){,\osa322,\osa634,\osa946,\osa{12}58,1
,$\gamma-\tfrac32\delta$}
\multiputlist(0,25)(40,0){$-2z$, ,\osa{11}66,\osa{16}98, ,2
,$\gamma-\tfrac52\delta$}
\multiputlist(0,0)(40,0){,,,\osa{23}{12}{12}, ,3
,$\gamma-\tfrac72\delta$}
\end{picture}
\caption{Observed degrees in $n$ (in square brackets) of various
solutions. The denominator always factorizes. For the $-2z$ hierarchy we
have subtracted $-2z$ from each solution.}
\end{figure}
{}From it we can see  that for the $-2z$ hierarchy the denominators
of $x_{MN}$ factorize with factors of degree $N(M+1)$ and $(N+1)M$ in
$n$, and for the $-1/z$ hierarchy the degrees are $N(M-N)$ and
$(N+1)(M-N)$.

Next observe that when $N=0$ one of the factors in the denominator is
constant and the other grows {\em linearly} with $M$. (For this value
$N=0$ the denominators are the same for the $-2z$ and $-1/z$
hierarchies.) The denominator is interpreted as the product of an
$0\times0$ and $1\times1$ matrix, and this suggests that the
denominators can provide us the basic polynomials that are to be used
as matrix elements. The first few polynomials obtained this way are
(when normalized to be monic):
\begin{eqnarray*}
p_0(n,c)&:=&1,\\ p_1(n,c)&:=&n,\\ p_2(n,c)&:=& n^2 - \tfrac12 c - 1,\\
p_3(n,c)&:=& n^3 + n ( - \tfrac32 c - 4),\\ p_4(n,c)&:=& n^4 + n^2 ( -
3 c - 10) + \tfrac34 c^2 + \tfrac{21}4 c + 9,\\ p_5(n,c)&:=& n^5 + 5
n^3 ( - c - 4) + n ( \tfrac{15}4 c^2 + \tfrac{125}4 c + 64),\\
p_6(n,c)&:=& n^6 + n^4 ( - \tfrac{15}2 c - 35) + n^2 (\tfrac{45}4 c^2
+ \tfrac{435}4 c + 259)\\ &&\quad - \tfrac{15}8 c^3 - \tfrac{225}8 c^2
- \tfrac{555}4 c - 225,\\ p_7(n,c)&:=& n^7 + n^5 ( - \tfrac{21}2 c -
56) + n^3 (\tfrac{105}4 c^2 + \tfrac{1155}4 c + 784)\\ &&\quad + n ( -
\tfrac{105}8 c^3 - \tfrac{1785}8 c^2 - \tfrac{2499}2 c - 2304)
\end{eqnarray*}

One indication that we are on the right track is obtained when we
observe that the $p_N(n,c)$ satisfy recursion relations:
\begin{equation}\begin{array}{lll}
&&p_{N+1}(n,c)=n\,p_{N}(n,c)-\tfrac{N}2(c+N+1)\,p_{N-1}(n,c+1),\\
&&p_N(n+\tfrac12,c)-p_N(n-\tfrac12,c)=N\,p_{N-1}(n,c+\tfrac12).
\end{array}\label{E:recdH}
\end{equation}
As was noted in Sec.~1 the matrix elements of the corresponding continuous 
case are given in terms of Hermite
polynomials. Now we note that in the continuous limit
(which means $n=x/(\sqrt2 \delta),\,c=1/\delta^2,\,\delta\to0$)
we get for
\begin{equation}
H_N(x):= \lim_{\delta\to0}\,(\sqrt2\delta)^{N}
p_N(x/(\sqrt2\delta),1/\delta^2),
\end{equation}
the recursions relations of Hermite polynomials (\ref{Hermite}).

\section{Matrix form of the denominators}
The above indicates what the polynomial matrix entries should be, and
determining the structure of these matrices in the denominator is the
next problem. 

For $N=0$ the denominator was interpreted as the product of an
$0\times0$ and $1\times1$ matrix.  For $N=1$ we can see from the table
that one of the factors grows by 1 as $M$ increases by 1 and the other
factor increases by 2.  These factors were interpreted as $1\times1$ and
$2\times2$ matrices, respectively.  The $1\times1$ part for $N=1$ was
then found to be proportional to the {\em shifted} basic polynomials
given above: for a given $M$ this factor is $p_{M+1}(c-1)$ in the $-2z$
hierarchy and $p_{M+1}(c+1)$ in the $-1/z$ hierarchy.  The necessity of
shifts in $c$ is an important new ingredient, and something that exists
only in the discrete version. 

The next problem was to find a suitable matrix structure for the
factors of the denominator. Our working assumption was that for any
$M$ the denominator should be a product of a $N\times N$ and a
$(N+1)\times(N+1)$ matrix.  The degrees of these factors led us to
try the determinantal structure of Hankel type
\[
T_{L,K}:=\left|\begin{array}{cccc}
q_K&q_{K+1}&\cdots&q_{K+L-1}\\ 
q_{K+1}&q_{K+2}&\cdots&q_{K+L}\\ 
\vdots&\vdots&\ddots&\vdots\\
q_{K+L-1}&q_{K+L}&\cdots&q_{K+2L-2}
\end{array}\right|,
\]
because now if the entries $q_K$ are non-monic polynomials of degree
$K$ then $deg(T_{L,K})=L(K+L-1)$.

A more detailed study with some trial and error revealed that $q$'s
are not simply proportional to the $p$'s obtained before, but that
different shifts in $c$ are also needed for the matrix elements. Our
final result is as follows:

For the $-2z$ hierarchy (for which $\alpha= \gamma+(M+\frac12)\delta,
\,\beta=\gamma-(N+\frac12)\delta$ where $0\le N\le M$) the denominator
can be expressed as
\begin{equation}
den(x_{M,N})=\tau_{N,M-N+2}(-1)\,\tau_{N+1,M-N}(0),
\end{equation}
where
\begin{eqnarray}
\hskip -0.6cm\tau_{L,K}(s):=&&\nonumber\\
&&\hskip -1.7cm\left|\begin{array}{cccc}
q_K(s)&q_{K+1}(s)&\cdots&q_{K+L-1}(s)\\ 
q_{K+1}(s-1)&q_{K+2}(s-1)&\cdots&q_{K+L}(s-1)\\ 
\vdots&\vdots&\ddots&\vdots\\
q_{K+L-1}(s-L+1)&q_{K+L}(s-L+1)&\cdots&q_{K+2L-2}(s-L+1)
\end{array}\right|,\label{E:tau}
\end{eqnarray}
and
\begin{equation}
q_M(s):=(c+s)^M\,p_M(c+s)/M!\,.
\end{equation}

For the $-1/z$ hierarchy ($\alpha=\gamma+(M+\frac12)\delta,\,
\beta=\gamma+(N+\frac12)\delta$, where $0\le N< M$) we found
same matrix form, but with different degrees and shifts:
\begin{equation}
den(x_{M,N})=\tau_{N,M-2N+1}(N)\,\tau_{N+1,M-2N}(N).
\end{equation}

\section{The numerator}
Finding a determinantal form for the numerator was more difficult,
because it did not factorize. However this was expected, because
usually it turns out that the numerator is the sum of two products of
$\tau$-functions, c.f., (\ref{E:taup2},\ref{-1/z},\ref{-2z}).

Thus we tried to express the numerator as a sum of two products of two
$\tau$-functions (\ref{E:tau}), with possible shifts not only in $c$,
but also in $n$, corresponding to the derivatives in the continuous
cases. Furthermore it seemed reasonable to assume that the sizes of
the matrices were the same as in the denominators, and that the other
index of the $\tau$-function depended linearly on $M$ and $N$.
Figuring out the shifts in $n$ required some more trial and error, but
eventually we arrived at the result that worked:

For the $-2z$ hierarchy we got
\begin{eqnarray}
&&\hskip -1cm
x(n;\gamma+(M+\tfrac12)\delta,\gamma-(N+\frac12)\delta,\gamma)
\propto\nonumber\\ &&\hskip 2cm\frac{\cosh(\tfrac12
D_n)\,\tau_{N,M-N+1}(-\tfrac12)\cdot \tau_{N+1,M-N+1}(-\tfrac12)}
{\tau{}_{N,M-N+2}(-1)\,\tau_{N+1,M-N}(0)},
\end{eqnarray}
and the corresponding result for the $-1/z$ hierarchy was
\begin{eqnarray}
&&\hskip -0.5cm x(n;\gamma+(M+\tfrac12)\delta,
\gamma+(N+\frac12)\delta,\gamma)
\propto\nonumber\\
&&\hskip 1.5cm\frac{
\sinh(\tfrac12 D_n)\, \tau{}_{N,M-2N+1}(N-\tfrac12)\cdot
\tau{}_{N+1,M-2N}(N+\tfrac12)} 
{\tau{}_{N,M-2N+1}(N)\,\tau_{N+1,M-2N}(N)}.
\end{eqnarray}
In expressing the $n$-shifts we have used the usual Hirota bilinear
derivative operator $D_n$. These results look a bit more symmetrical if
we define $\bar \tau_{N,A+B+N-1}(2B-N+1):=\tau_{N,A}(B)$, then we get
\begin{eqnarray}
&&\hskip -1cm
x(n;\gamma+(M+\tfrac12)\delta,\gamma-(N+\frac12)\delta,\gamma)
\propto\nonumber\\ &&\hskip 2cm\frac{\cosh(\tfrac12 D_n)\,\bar
\tau_{N,M-\frac12}(-N)\cdot \bar \tau_{N+1,M+\frac12}(-N-1)} {\bar
\tau{}_{N,M}(-N-1)\,\bar \tau_{N+1,M}(-N)},
\end{eqnarray}
and
\begin{eqnarray}
&&\hskip -0.5cm x(n;\gamma+(M+\tfrac12)\delta,
\gamma+(N+\frac12)\delta,\gamma)
\propto\nonumber\\
&&\hskip 1.5cm\frac{
\sinh(\tfrac12 D_n)\, \bar \tau{}_{N,M-\frac12}(N)\cdot
\bar \tau{}_{N+1,M+\frac12}(N+1)} 
{\bar \tau{}_{N,M}(N+1)\,\bar \tau_{N+1,M}(N)}.
\end{eqnarray}

\pagebreak

\section{Conclusions} We have shown here that the two hierarchies of
rational solutions for d-\Pf\ (generated from $-2z$ and $-1/z$) can be
expressed in terms of determinants, and that the matrix elements of
these determinants are given by polynomials that can be regarded as
discrete analogues of Hermite polynomials.  It is interesting that in
these expressions the parameter $c$, which vanishes in the continuous
limit, plays an important role.  However, the result is not yet
complete, because the bilinear equation for the $\tau$-function is still
to be written, and the general proof must be given. 

For the $-\frac23 z$ hierarchy things are an order of magnitude more
difficult. There is no linear growth in any direction in the parameter
space so there are no candidates for matrix elements. We have
nevertheless found what the $\tau$-functions should be, but no
determinantal expression for them. Although no determinantal
expression or basic polynomial is known even in the continuous case,
we hope that similar structures in solutions of both continuous and
discrete cases will be found.

\section*{Acknowledgments}
One of the authors (K.K) was supported by the Grant-in-Aid for
Encouragement of Young Scientists from The Ministry of Education,
Science, Sports and Culture of Japan, No.08750090. This work was
started when J.H. was visiting Doshisha University on an exchange
grant between the Japanese Society for the Promotion of Science and
the Academy of Finland.


\begin{thebibliography}{99}
\bibitem{Luka} N.~Lukashevich, Diff.~Urav.~{\bf 1}, 731 (1965). [Diff.
Eqs. {\bf 1}, 561 (1965).] N.~Lukashevich, Diff.~Urav.~{\bf 3}, 771
(1967). [Diff.  Eqs. {\bf 3}, 395 (1967).] V.I.~Gromak,
Diff.~Urav.~{\bf 23}, 760 (1987). [Diff.~ Eqs. {\bf 23}, 506 (1987).]
\bibitem{Oka}K.~Okamoto, Math. Ann. {\bf 275}, 221 (1986).
\bibitem{Murata} Y.~Murata, Funkcial. Ekvac., {\bf 28}, 1 (1985).
\bibitem{GRP} B. Grammaticos, A. Ramani and V. Papageorgiou,
Phys. Rev. Lett. {\bf 67}, 1825 (1991).
\bibitem{RGH} A. Ramani, B. Grammaticos and J. Hietarinta,
Phys. Rev. Lett. {\bf 67}, 1829 (1991).
\bibitem{coales} A. Ramani and B. Grammaticos,
 Physica {\bf A 228}, 160 (1996).
\bibitem{bi} A. Ramani, B. Grammaticos and J. Satsuma: 
J. Phys. A: Math. Gen. {\bf 28}, 4655 (1995).
\bibitem{Vorobev} A.P. Vorob'ev, Diff.~Urav.~{\bf 1}, 79 (1965). 
[Diff. Eqs. {\bf 1}, 58 (1965).]
\bibitem{P2} K. Kajiwara and Y. Ohta, J. Math. Phys. {\bf 37},
4693 (1996).
\bibitem{dP2-rational_1} J. Satsuma, K. Kajiwara, B. Grammaticos, 
J. Hietarinta and A. Ramani, J. Phys. A: Math.~Gen., 
{\bf 28} 3541 (1995).
\bibitem{dP2-rational_2} K. Kajiwara, K. Yamamoto and Y. Ohta,
preprint {\tt solv-int/9702001}.
\bibitem{KOu} K. Kajiwara and Y. Ohta, unpublished
\bibitem{Schl} K.M. Tamizhmani, B. Grammaticos and A. Ramani,
Lett. Math. Phys. {\bf 29}, 49 (1993).
\bibitem{CB} A. Bassom and P. Clarkson, Phys. Lett. A {\bf 194} 358
(1994); P. Clarkson and A. Bassom, preprint {\tt solv-int/9412002}.
\bibitem{RED}  A.C. Hearn, {\it REDUCE User's Manual}, v.3.6
\copyright 
 Rand (1995).
\end{thebibliography}
\end{document}